\documentclass[aps,prd,preprint,groupedaddress,showpacs]{revtex4}

\usepackage{epsfig}
\usepackage{graphics}
\usepackage{slashed}
\usepackage{amssymb,amsmath}

\begin{document}

\leftmargin -2cm
\def\choosen{\atopwithdelims..}



 \boldmath
\title{Dijet azimuthal decorrelations at the LHC\\ in the parton Reggeization approach } \unboldmath

\author{\firstname{M.A. }\surname{Nefedov}}
\email{nefedovma@gmail.com}
\affiliation{Samara State University, Ac. Pavlov, 1, 443011 Samara,
Russia}
\author{\firstname{V.A.} \surname{Saleev}} \email{saleev@samsu.ru}

\author{\firstname{A.V.} \surname{Shipilova} }\email{alexshipilova@samsu.ru}
\affiliation{Samara State University, Ac. Pavlov, 1, 443011 Samara,
Russia}
\affiliation{{II.} Institut f\"ur Theoretische Physik, Universit\" at Hamburg,
Luruper Chaussee 149, 22761 Hamburg, Germany}

\begin{abstract}
We study inclusive dijet azimuthal decorrelations in proton-proton
collisions at the CERN LHC invoking the hypothesis of parton
Reggeization in $t-$channel exchanges at high energies.
 In the parton Reggeization
approach, the main contribution to the azimuthal angle difference
between the two most energetic jets is due to the
Reggeon-Reggeon-particle-particle scattering, when the fusion of two
Reggeized gluons into a pair of Yang-Mills gluons dominates. Using a
high-energy factorization scheme with the Kimber-Martin-Ryskin
unintegrated parton distribution functions and the Fadin-Lipatov
effective vertices we obtain good agreement of our calculations with
recent measurements by the ATLAS and CMS Collaborations at the CERN
LHC.
\end{abstract}

\pacs{12.38.Bx, 12.39.St, 13.87.Ce}
\maketitle

\section{Introduction}
\label{sec:one}

The production of large transverse-momentum ($p_T$) hadron jets at
the high-energy colliders, like the Tevatron and the LHC, is
reasonably considered as an important source of information about
the dynamics of parton-parton interactions at small distances and
parton distribution functions (PDF), and as a probe to test
perturbative quantum chromodynamics (pQCD) \cite{Review}. The
measurements of decorrelations in the azimuthal angle between the
two most energetic jets, $\Delta\varphi$, as a function of the
number of produced jets, give the chance to separate directly
leading-order (LO) and next-to-leading orders (NLO) contributions in
the strong coupling constant $\alpha_s$. Furthermore, a precise
understanding of the physics of events with large azimuthal
decorrelations is essential for the search for new physical
phenomena with dijet signatures by the CMS \cite{CMS2} and ATLAS
\cite{ATLAS2} detectors at the LHC.

The total collision energies, $\sqrt{S}=7$ TeV or 14 TeV at the LHC,
sufficiently exceed the transverse momenta of identified jets
($0.1<p_T<1.3$ TeV). The recent theoretical studies
 of single-jet  production at very high energy show the
dominance of the multi-Regge final states in the inclusive
production cross sections \cite{KSSY}. In other words, in the
high-energy regime the contribution of partonic subprocesses
involving $t-$channel parton exchanges become dominant. Then the
transverse momenta of the incoming partons and their off-shell
properties can no longer be neglected, and we deal with
QCD-scattering amplitudes with Reggeized $t-$channel partons. As it
has been shown by Lipatov and co-authors, Reggeized gluons and
quarks are the appropriate gauge-invariant degrees of freedom of
high-energy pQCD. In the practical applications, the use of
Reggeized $t-$channel exchanges is justified by the dominance of the
so-called multi-Regge final states in inelastic collisions of high-energy hadrons.

The parton Reggeization approach \cite{FadinLipatov,FadinSherman} is
based on an effective quantum field theory implemented with the
non-Abelian gauge-invariant   action which includes fields of
Reggeized gluons \cite{Lipatov95} and Reggeized quarks
\cite{LipatoVyazovsky}. Recently, this approach was successfully
applied to interpret the $p_T-$spectra of inclusive production of a
single jet \cite{KSSY}, prompt photon \cite{tevatronY,heraY},
Drell-Yan lepton pairs \cite{NNS_DY} and bottom-flavored jets
\cite{bbTEV,bbLHC} at the Tevatron and LHC. The single-jet
production is  dominated by the multi-Regge kinematics (MRK),  when
only a jet with a highest transverse momentum (leading jet) is
produced in the central  rapidity region, being strongly separated
in rapidity from other particles. If the same situation is realized
for two or more the most energetic jets, then quasimulti-Regge
kinematics (QMRK) is at work.

 In the present paper we extend our analysis of inclusive single-jet
production at the LHC  based on the parton Reggeization approach to
a case of dijet production, and study azimuthal decorrelations
between the two central jets with the largest transverse momenta
according to the measurements implemented by the CMS and ATLAS
Collaborations \cite{CMS2,ATLAS2}.

 This paper is organized as follows: In Sec.~II, the parton
Reggeization approach is briefly reviewed. We write down relevant
Reggeon-Reggeon-particle-particle effective vertices, effective
amplitudes, squared matrix elements and differential cross sections.
In Sec.~III, we present the results of our calculations for two-jet
azimuthal decorrelation spectra and perform a comparison with
corresponding experimental data from the CERN LHC at $\sqrt{S}=7$
TeV. Sec.~IV contains our conclusions.

\section{Dijet production in QMRK}
\label{sec:two}

In the high-energy Regge limit, when the collision energy is very
high but the transverse momenta of the produced jets are fixed by
the condition $\sqrt{S}>>p_T>>\Lambda_{QCD}$, the high-energy
(uncollinear) factorization works well instead of collinear parton
model. In this case, the Balitsky-Fadin-Kuraev-Lipatov (BFKL) QCD
evolution of unintegrated PDFs \cite{BFKL} can be more adequate
approximation than the Dokshtser-Gribov-Lipatov-Altarelli-Parisi
(DGLAP) QCD evolution of collinear PDFs \cite{DGLAP}. This means,
that we need to consider the processes with strong ordering in
rapidity at the MRK or QMRK conditions. We  identify final-state
jets and final-state partons, and consider the production of two
partons in the central region of rapidity, assuming that there are
no any other partons with the same rapidity. To the LO in the parton
Reggeization approach we have the following partonic subprocesses,
which describe the production of two central jets in proton-proton
collisions:
\begin{eqnarray}
R+R \to g+g,\label{RRgg}\\
R+R \to q+ \bar q,\label{RRqq}\\
Q+R \to q+g,\label{RQgq}\\
Q+Q\to q+q,\label{QQqq1}\\
Q+Q'\to q+q',\label{QQqq2}\\
 Q+\bar Q\to q+\bar q,\label{QaQqaq1}\\
 Q+\bar Q\to q'+\bar q',\label{QaQqaq2}\\
 Q+\bar Q\to g+g,\label{QaQgg}
\end{eqnarray}
where $R$ is a Reggeized gluon, $Q$ is a Reggeized quark, $g$ is a
Yang-Mills gluon, $q$ is an ordinary quark, $q$ and $q'$ denote
quarks of different flavors. Working in the center-of-mass (c.m.)
frame, we write the four-momenta of the incoming protons as
$P_{1,2}^\mu=(\sqrt{S}/2)(1,0,0,\pm 1)$ and those of the Reggeized
partons as $q_i^\mu=x_i P_i^\mu+q_{iT}^\mu\ (i=1,2)$, where $x_i$ are
 longitudinal momentum fractions and $q_{iT}^\mu=(0,{\bf
q}_{iT},0)$, with ${\bf q}_{iT}$ being transverse two-momenta, and
we define $t_i=-q_{iT}^2={\bf q}_{iT}^2$. The final partons have
four-momenta $k_{1,2}$ and they are on-shell and massless
$k_1^2=k_2^2=0$.

The effective gauge-invariant amplitudes for the above-mentioned
subprocesses (\ref{RRgg})-(\ref{QaQgg}) can be obtained using the
effective Feynman rules from Refs.~
\cite{Lipatov95,LipatoVyazovsky,Antonov}. As usual, we introduce the
light-cone vectors $n^{+}=2P_2/\sqrt{S}$ and $n^{-}=2P_1/\sqrt{S}$,
and define $k^{\pm}=k\cdot n^{\pm}$ for any four-vector $k^\mu$.
Than we determine effective vertices:
\begin{eqnarray}
\Gamma^{(+-)}_\mu(q_1,q_2)&=&2\left[\left(q_1^++\frac{q_1^2}{q_2^-}\right)n^-_\mu-\left(q_2^-+\frac{q_2^2}{q_1^+}\right)n^+_\mu+(q_2-q_1)_\mu\right],\\
\gamma_\mu^{\pm}(q,p)&=&\gamma_\mu+\hat q \frac{n^\pm_\mu}{p^\pm},\\
\gamma_\mu^{(+-)}(q_1,q_2)&=& \gamma_\mu-\hat
q_1\frac{n_\mu^-}{q_2^-}-\hat q_2\frac{n_\mu^+}{q_1^+},
\end{eqnarray}
\begin{eqnarray}
\Gamma^{\mu\nu +}(q_1,q_2)&=&2q_1^+ g^{\mu\nu}-(n^+)^\mu(q_1-q_2)^\nu-(n^+)^\nu(q_1+2q_2)^\mu+
\frac{t_2}{q_1^+}(n^+)^\mu(n^+)^\nu,\nonumber\\
\Gamma^{\mu\nu -}(q_1,q_2)&=&2q_2^- g^{\mu\nu}+(n^-)^\mu(q_1-q_2)^\nu-(n^-)^\nu(q_2+2q_1)^\mu+
\frac{t_1}{q_2^-}(n^-)^\mu(n^-)^\nu,
\end{eqnarray}
and the triple-gluon vertex
\begin{eqnarray}
\gamma_{\mu\nu\sigma}(q,p)&=&(q-p)_\sigma g_{\mu\nu}-(p+2 q)_\mu
g_{\nu\sigma}+(2p+q)_\nu g_{\mu\sigma}.
\end{eqnarray}
Finally,
 we can write the effective vertices for relevant subprocesses as
 follows:
\begin{eqnarray}
C_{RR,ab}^{gg,~cd,~\mu\nu}(q_1,q_2,k_1,k_2)&=&g_s^2\frac{q_1^+q_2^-}{4\sqrt{t_1t_2}}\biggl( T_1
s^{-1}\Gamma^{(+-)_\sigma}(q_1,q_2)\gamma_{\mu\nu\sigma}(-k_1,-k_2)+\nonumber\\
&+&T_3 t^{-1}\Gamma^{\sigma\mu -}(q_1,k_1-q_1)\Gamma^{\sigma\nu
+}(k_2-q_2,q_2)-\nonumber\\
&-&T_2 u^{-1}\Gamma^{\sigma\nu -}(q_1,k_2-q_1)\Gamma^{\sigma\mu
+}(k_1-q_2,q_2)-\nonumber\\
&-&T_1\bigl(n^-_\mu n^+_\nu- n^-_\nu n^+_\mu
\bigr)-T_2\bigl(2g_{\mu\nu}- n^-_\mu n^+_\nu
\bigr)-T_3\bigl(-2g_{\mu\nu}+ n^-_\nu n^+_\mu
\bigr)+\nonumber\\
&+&\Delta^{\mu\nu+}(q_1,q_2,k_1,k_2)+\Delta^{\mu\nu-}(q_1,q_2,k_1,k_2)\biggr),\label{verRRgg}
\end{eqnarray}
where
\begin{eqnarray}
&&T_1=f_{cdr}f_{abr}, \quad T_2=f_{dar}f_{cbr}, \quad
T_3=f_{acr}f_{dbr},\quad T_1+T_2+T_3=0,\nonumber\\
&&\Delta^{\mu\nu+}_{}(q_1,q_2,k_1,k_2)=2t_2n^+_\mu
n^+_\nu\biggl(
\frac{T_3}{k_2^+ q_1^+}-\frac{T_2}{k_1^+ q_1^+}\biggr),\nonumber\\
&&\Delta^{\mu\nu-}_{}(q_1,q_2,k_1,k_2)=2t_1 n^-_\mu
n^-_\nu\biggl( \frac{T_3}{k_1^- q_2^-}-\frac{T_2}{k_2^-
q_2^-}\biggr),\nonumber
\end{eqnarray}
 $f^{abc}$ are structure constants of the color gauge group SU(3),
$g_s^2=4 \pi \alpha_s$, and $\alpha_s$ is a strong coupling
constant.


\begin{eqnarray}
C_{RR,~ab}^{q\bar q}(q_1,q_2,k_1,k_2)&=&g_s^2\frac{q_1^+q_2^-}{4\sqrt{t_1t_2}}\bar U(k_1)\left( -s^{-1} [T^a, T^b]\gamma^\sigma \Gamma^{(+-)}_\sigma(q_1,q_2)+\right.\nonumber\\
&+& \left.t^{-1}T^a T^b \gamma^-(\hat k_1-\hat q_1)\gamma^+ +  u^{-1}T^b T^a \gamma^+(\hat k_1-\hat q_2)\gamma^-\right)V(k_2),\label{verRRqq}
\end{eqnarray}
\begin{eqnarray}
C_{QR,~a}^{qg,~b,~\mu}(q_1,q_2,k_1,k_2)&=&\frac{1}{2}g_s^2\frac{q_2^-}{2\sqrt{t_2}}\bar
U(k_1)\biggl[ \gamma_\sigma^{(-)}(q_1,k_1-q_1)t^{-1}\bigl(
\gamma_{\mu\nu\sigma}(k_2,-q_2)n_\nu^+ +  t_2\frac{n_\mu^+
n_\sigma^+}{k_2^+}\bigr)  \times \nonumber\\
&\times &\bigl[T^a,T^b\bigr]-\gamma^+(\hat q_1-\hat
k_2)^{-1}\gamma_\mu^{(-)}(q_1,-k_2)T^aT^b-\nonumber\\
&-&\gamma_\mu(\hat q_1+\hat
q_2)^{-1}\gamma_\sigma^{(-)}(q_1,q_2)n_\sigma^+T^bT^a+\frac{2\hat
q_1 n_\mu^-}{k_1^-}\biggl(\frac{T^aT^b}{k_2^-}-
\frac{T^bT^a}{q_2^-}\biggr)\biggr],\label{verRQgq}
\end{eqnarray}
\begin{eqnarray}
C_{QQ}^{qq}(q_1,q_2,k_1,k_2)&=&g_s^2\left(t^{-1} \bar U(k_2)
\gamma_\sigma^{(+)}(q_2,k_2-q_2)T^c\otimes \bar
U(k_1)\gamma_{\sigma}^{(-)}(q_1, k_1-q_1)T^c-\right.\nonumber\\
&-&\left.u^{-1} \bar U(k_1)
\gamma_\sigma^{(+)}(q_2,k_1-q_2)T^c\otimes \bar
U(k_2)\gamma_{\sigma}^{(-)}(q_1,k_2-q_1)T^c\right),\label{verQQqq1}
\end{eqnarray}
\begin{eqnarray}
C_{QQ'}^{qq'}(q_1,q_2,k_1,k_2)=g_s^2t^{-1} \bar U(k_2)
\gamma_\sigma^{(+)}(q_2,k_2-q_2)T^c\otimes \bar
U(k_1)\gamma_{\sigma}^{(-)}(q_1, k_1-q_1)T^c,\label{verQQqq2}
\end{eqnarray}
\begin{eqnarray}
C_{Q\bar Q}^{q\bar q}(q_1,q_2,k_1,k_2)&=&g_s^2\left(s^{-1}\bar
U(k_1)\gamma_{\sigma}T^c V(k_2) \otimes
\gamma_{\sigma}^{(+-)}(q_1,q_2) T^c+\right.\nonumber\\
&+&\left.t^{-1}\bar U(k_1)
\gamma_\sigma^{(+)}(q_2,k_2-q_2)T^c\otimes
\gamma_\sigma^{(-)}(q_1,k_1-q_1)T^c V(k_2)\right),\label{verQaQqaq1}
\end{eqnarray}
\begin{eqnarray}
C_{Q\bar Q}^{q'\bar q'}(q_1,q_2,k_1,k_2)=g_s^2s^{-1}\bar
U(k_1)\gamma_{\sigma}T^c V(k_2) \otimes
\gamma_{\sigma}^{(+-)}(q_1,q_2) T^c,\label{verQaQqaq2}
\end{eqnarray}
\begin{eqnarray}
C_{Q\bar
Q}^{gg,~ab,~\mu\nu}(q_1,q_2,k_1,k_2)&=&-g_s^2\biggl(\gamma_\nu^{(+)}(q_2,-k_2)(\hat
k_2-\hat
q_2)^{-1}\gamma_\mu^{(-)}(q_1,-k_1)T^bT^a+\nonumber\\
&+& \gamma_\nu^{(+)}(q_2,-k_1)(\hat k_1-\hat
q_2)^{-1}\gamma_\mu^{(-)}(q_1,-k_2)T^aT^b+\nonumber\\
&+&\gamma_{\mu\nu\sigma}(-k_1,-k_2)s^{-1}\gamma^{(+-)}_{\sigma}(q_1,q_2)[T^a,T^b]+
\Delta_{\mu\nu}^{ab}(q_1,q_2)\biggr),\label{verQaQgg}
\end{eqnarray}
where
$$\Delta_{\mu\nu}^{ab}(q_1,q_2)=\frac{\hat q_1n_\mu^-
n_\nu^-}{q_2^-}\biggl(
\frac{T^aT^b}{k_2^-}+\frac{T^bT^a}{k_1^-}\biggr)-\frac{\hat q_2n_\mu^+ n_\nu^+}{q_1^+}\biggl( \frac{T^aT^b}{k_1^+}+
\frac{T^bT^a}{k_2^+}\biggr).$$
 The effective amplitudes of the subprocesses
(\ref{RRgg})-(\ref{QaQgg}) are straightforwardly found from Eqs.
(\ref{verRRgg})-(\ref{verQaQgg}). In the case of gluon production
amplitudes we convolute effective vertices with final gluon
polarization vectors $\epsilon_{\mu}^a(k_{1})$ and
$\epsilon_{\nu}^b(k_{2})$. In case of amplitudes with the
initial-state Reggeized quark $Q$ and antiquark $\bar Q$, we take
their spinors  as $U(x_1P_1)$ and $V(x_2P_2)$
\cite{LipatoVyazovsky}. In such a way, omitting the color indices,
we obtain
\begin{eqnarray}
{\cal M}(RR\to gg)=\epsilon_{\mu}(k_1)\epsilon_{\nu}(k_2)
C_{RR}^{gg,~\mu\nu}(q_1,q_2,k_1,k_2),\label{ampRRgg}
\end{eqnarray}
\begin{eqnarray}
{\cal M}(RR\to q\bar q)=C_{RR}^{q\bar
q}(q_1,q_2,k_1,k_2),\label{ampRRqq}
\end{eqnarray}
\begin{eqnarray}
{\cal M}(QR\to qg)=\epsilon_{\mu}(k_2)
C_{QR}^{qg,~\mu}(q_1,q_2,k_1,k_2)U(x_1P_1),\label{ampRQgq}
\end{eqnarray}
\begin{eqnarray}
{\cal M}(QQ\to qq)=C_{QQ}^{qq}(q_1,q_2,k_1,k_2)U(x_1P_1)\otimes
U(x_2P_2),\label{ampQQqq1}
\end{eqnarray}
\begin{eqnarray}
{\cal M}(QQ'\to qq')=C_{QQ'}^{qq'}(q_1,q_2,k_1,k_2)U(x_1P_1)\otimes
U'(x_2P_2),\label{ampQQqq2}
\end{eqnarray}
\begin{eqnarray}
{\cal M}(Q\bar Q\to q\bar q)=\bar V(x_2P_2)C_{Q\bar Q}^{q\bar
q}(q_1,q_2,k_1,k_2)U(x_1P_1),\label{ampQaQqaq1}
\end{eqnarray}
\begin{eqnarray}
{\cal M}(Q\bar Q\to q'\bar q')=\bar V(x_2P_2) C_{Q\bar Q}^{q'\bar
q'}(q_1,q_2,k_1,k_2)U(x_1P_1),\label{ampQaQqaq2}
\end{eqnarray}
\begin{eqnarray}
{\cal M}(Q\bar Q\to gg)= \epsilon_{\mu}(k_1)\epsilon_{\nu}(k_2)\bar
V(x_2P_2) C_{Q\bar
Q}^{gg,~\mu\nu}(q_1,q_2,k_1,k_2)U(x_1P_1).\label{ampQaQgg}
\end{eqnarray}

To calculate dijet production cross sections we have found squared
amplitudes $\overline{|{\cal M}|^2}$ of the above-mentioned
subprocesses (\ref{RRgg})-(\ref{QaQgg}), where the bar indicates
average (summation) over initial-state (final-state) spins and
colors. The results are written as functions of the Mandelstam
variables $s=(q_1+q_2)^2$, $t=(q_1-k_1)^2$, $u=(q_1-k_2)^2$, and
invariant Sudakov variables $a_1=2 k_1\cdot P_2/S$, $a_2=2 k_2\cdot
P_2/S$, $b_1=2 k_1\cdot P_1/S$, $b_2=2 k_2\cdot P_1/S$. The
four-momenta of final-state partons can be introduced as a sum of
longitudinal and transverse parts: $k_1=a_1 P_1+b_1 P_2+k_{1T}$,
$k_2=a_2 P_1+b_2 P_2+k_{2T}$. Applying a four-momentum conservation
law one can find $x_1=a_1+a_2$, $x_2=b_1+b_2$ and
$q_{1T}+q_{2T}=k_{1T}+k_{2T}$. In the general case, the squared
amplitudes can be written in the form
\begin{equation}
\overline{|{\cal M}|^2}=\pi^2 \alpha_S^2 A \sum_{n=0}^4 W_n S^n,
\end{equation}
where $A$ and $W_n$ are process-dependent functions of variables
$s,t,u,a_1,a_2,b_1,b_2,t_1,t_2,S$. The exact analytical
 formulas for $A$ and $W_n$ are presented in  the Appendix. Our
definition of the Reggeized amplitudes satisfies  evident the
normalization to the QCD amplitudes of the collinear parton model:
\begin{equation}
\lim_{t_1,t_2\to 0}\int \frac{d\varphi_1}{2\pi}\int
\frac{d\varphi_2}{2\pi}\overline{|{\cal
M}(t_1,t_2,\varphi_1,\varphi_2)|^2}=\overline{|{\cal M}_{PM}|^2}.
\end{equation}
 This condition has been checked for all squared matrix elements
$\overline{|{\cal M}|^2}$ presented in  the Appendix.

Here we present an analytic formula for the differential cross
section for the dijet production process via the subprocess
(\ref{RRgg}), which gives a main contribution, and take in mind that
other contributions can be written in the same manner. So, according
to the high-energy factorization formalism, the proton-proton
production cross section is obtained by the convolution of squared
matrix element $\overline{|{\cal M}(RR\to gg)|^2}$ with the
unintegrated Reggeized gluon PDFs
$\Phi^p_{g}(x_{1,2},t_{1,2},\mu^2)$ at the factorization scale $\mu$
\cite{tevatronYY}:
\begin{eqnarray}
\frac{d\sigma(pp\to gg
X)}{dk_{1T}dy_1dk_{2T}dy_2d\Delta\varphi}=\frac{k_{1T}k_{2T}}{16\pi^3}\int
dt_1\int d\phi_1 \Phi^p_{g}(x_1,t_1,\mu^2)\Phi^{
p}_{g}(x_2,t_2,\mu^2) \frac{\overline{|{\cal M}(RR\to
gg)|^2}}{(x_1x_2S)^2},\label{equSigma}
\end{eqnarray}
where $k_{1,2T}$ and $y_{1,2}$ are final gluon transverse momenta
and rapidities, respectively, and $\Delta\varphi$ is an azimuthal
angle enclosed between the vectors $\vec k_{1T}$ and $\vec k_{2T}$,
$$x_1=(k_1^0+k_2^0+k_1^z+k_2^z)/\sqrt{S}, \quad
x_2=(k_1^0+k_2^0-k_1^z-k_2^z)/\sqrt{S},$$
$$k_{1,2}^0=k_{1,2T}\cosh(y_{1,2}),
\quad k_{1,2}^z=k_{1,2T}\sinh(y_{1,2}).$$
Throughout our analysis the renormalization and factorization scales
are  identified and chosen to be  $\mu=\xi k_{1T}$, where $k_{1T}$ is
 the transverse momentum of  {the} leading jet ($k_{1T}>k_{2T}$) and $\xi$ is
varied between 1/2 and 2 about its default value 1 to estimate the
theoretical uncertainty due to the freedom in the choice of scales.
The resulting errors are indicated as shaded bands in the figures.

The unintegrated PDFs $\Phi_g^p(x,t,\mu^2)$ are related to their
collinear counterparts $f_g^p(x,\mu^2)$ by the normalization
condition
\begin{equation}
xf_g^p(x,\mu^2)=\int^{\mu^2}dt\,\Phi_g^p(x,t,\mu^2),
\end{equation}
which furnishes a correct transition from formulas in the parton
Reggeization approach to those in the collinear parton model.  In
our numerical analysis, we adopt the prescription proposed by
Kimber, Martin, and Ryskin (KMR) \cite{KMR} to obtain the
unintegrated gluon and quark PDFs of the proton from the
conventional integrated ones. As input for these procedures, we use
the LO set of the Martin-Roberts-Stirling-Thorne (MRST)
\cite{MRST2006} proton PDFs as our default.

\section{Results}

Recently, CMS \cite{CMS2}  and ATLAS \cite{ATLAS2} Collaborations
have measured azimuthal decorrelations between the two central jets
with the highest transverse momenta (leading jets) in proton-proton
collisions at $\sqrt{S}=7$ TeV. The kinematic domains of the jet's
transverse momenta and rapidities in these experiments are slightly
different. The CMS Collaboration selected two leading jets each with
$p_T>30$ GeV and rapidity $|y|<1.1$ and collected data into five
mutually exclusive regions, which are based on the $p_T^{max}$ in
the event: $80<p_T^{max}<110$ GeV, $110<p_T^{max}<140$ GeV,
$140<p_T^{max}<200$ GeV, $200<p_T^{max}<300$ GeV, and
$p_T^{max}>300$ GeV. The ATLAS Collaboration selected two leading
jets each with $p_T>100$ GeV and rapidity $|y|<0.8$ and collected
data into nine regions: $100<p_T^{max}<160$ GeV, $160<p_T^{max}<210$
GeV, $210<p_T^{max}<260$ GeV, $260<p_T^{max}<310$ GeV,
$310<p_T^{max}<400$ GeV, $400<p_T^{max}<500$ GeV,
$500<p_T^{max}<600$ GeV, $600<p_T^{max}<800$ GeV, and
$p_T^{max}>800$ GeV. The measurements are presented for the region
of ${\pi}/{2}<\Delta\varphi <\pi$ as normalized distributions
\begin{equation}F(\Delta\varphi)=\frac{1}{\sigma}\times \left(\frac{d\sigma}{d \Delta\varphi}\right),\end{equation}
 where
$$\sigma=\int_{\pi/2}^\pi \left(\frac{d\sigma}{d\Delta\varphi}\right)d\Delta\varphi.$$
Additionally, the ATLAS Collaboration presented the $\Delta\varphi$
distribution of events with $\geq 2$, $\geq 3$, $\geq 4$, and $\geq 5$ jets
with $p_T>100$ GeV and $|y|<0.8$  for the leading jets and $|y|<2.8$
for all other jets.

The theoretical  expectations based on the collinear parton model
for $\Delta\varphi$ distributions include pQCD calculations in NLO
($\alpha_S^4$) for the three-parton final states and LO
($\alpha_S^4$) for the four-parton final states \cite{NLO2jet}. As
it has been demonstrated in Refs.~ \cite{CMS2,ATLAS2}, these
calculations describe data in the region of $2 \pi/3 <\Delta\varphi<
\pi$ and overestimate data at $\Delta\varphi<2 \pi/3$. The agreement
of parton model calculations with data can be achieved using the
Monte Carlo event generators (MC), such as: PYTHIA \cite{pythia},
HERWIG++ and MADGRAPH \cite{herwig}. These parton-level generators
include NLO pQCD matrix elements, different collinear PDFs, effects
of hadronization and the initial-state parton shower radiation
(ISR). The last one is very important to simulate events at
$\Delta\varphi< 2 \pi/3$ but it is also needed to introduce a new
theoretically unknown parameter $k_{ISR}$, which can only be fixed
phenomenologically only.

To obtain a distribution $F(\Delta\phi)$ in the framework of the
parton Reggeization approach, we perform an integration of the
differential cross section (\ref{equSigma}) over the final-state
parton transverse momenta $k_{1T}$ and $k_{2T}$, as well as over the
rapidities $y_1$ and $y_2$ in the intervals defined by the
experiment. We take into account contributions of all necessary
subprocesses (\ref{RRgg})-(\ref{QaQgg}), where quark flavors are
taken  $q=u,d,s$ for initial-state and final-state quarks.  The
upper limit for the squares of the Reggeized gluon's transverse
momenta $t_1$ and $t_2$ should be truncated by the condition
$t_1,t_2<k_{2T}^2$, where $k_{2T}$ is a smaller transverse momentum
of a jet from the pair of two leading jets.

 The above-mentioned condition
arises from the constraints of the LHC experiment: one can measure
an azimuthal angle between the two most energetic jets
$\Delta\varphi$ but it is impossible to separate final-state partons
produced in the hard parton scattering phase from the ones generated
during QCD evolution of PDFs. The BFKL evolution suggests a strong
ordering in rapidity but transverse momenta of partons in the QCD
ladder keep similar values. This means, that  the transverse momenta
of partons
 generated in the initial-state evolution, described via the unintegrated
PDF must be smaller than transverse momenta of both
measured leading jets.

\newpage

\begin{figure}[ph]
\begin{center}
\includegraphics[width=.5\textwidth, clip=]{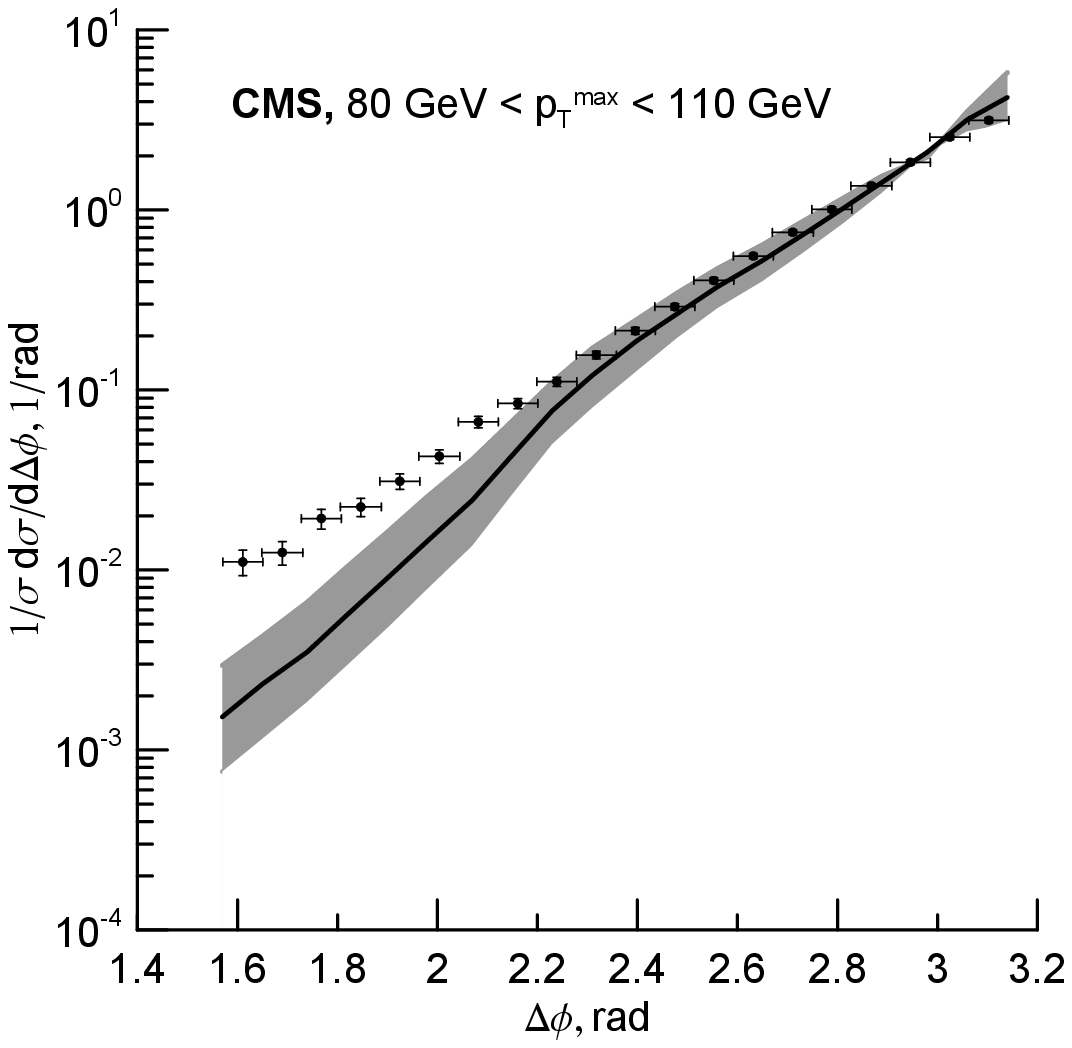}\includegraphics[width=.5\textwidth, clip=]{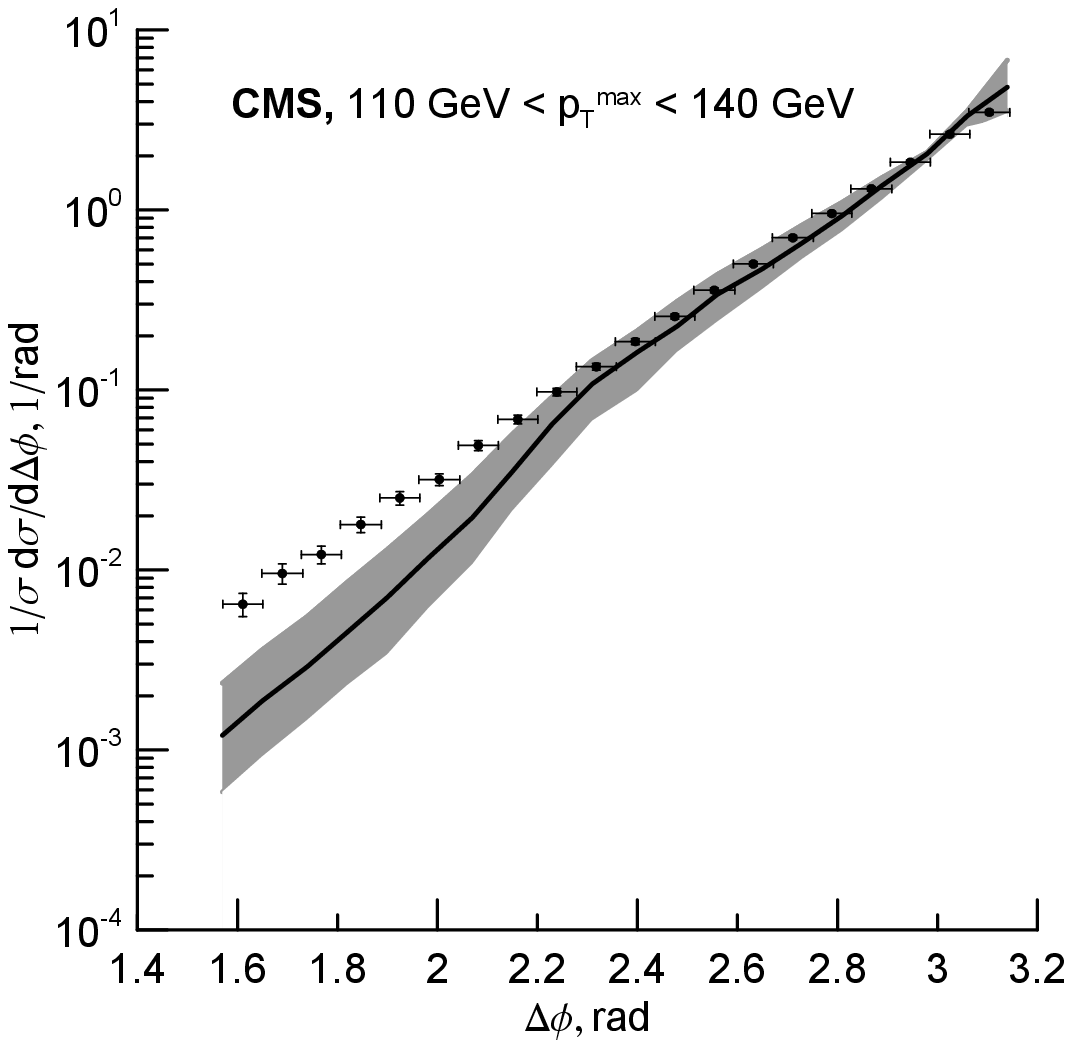}
\includegraphics[width=.5\textwidth, clip=]{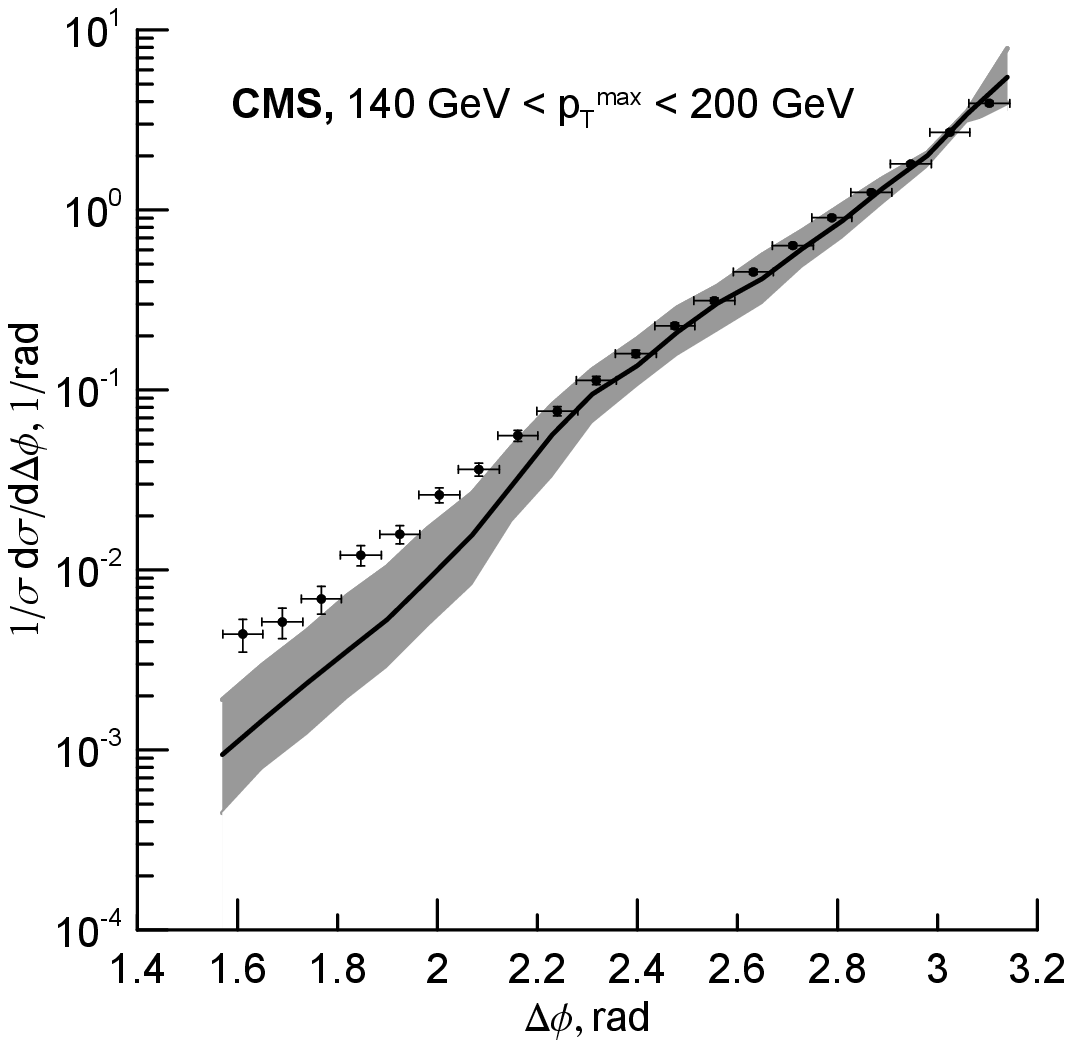}\includegraphics[width=.5\textwidth, clip=]{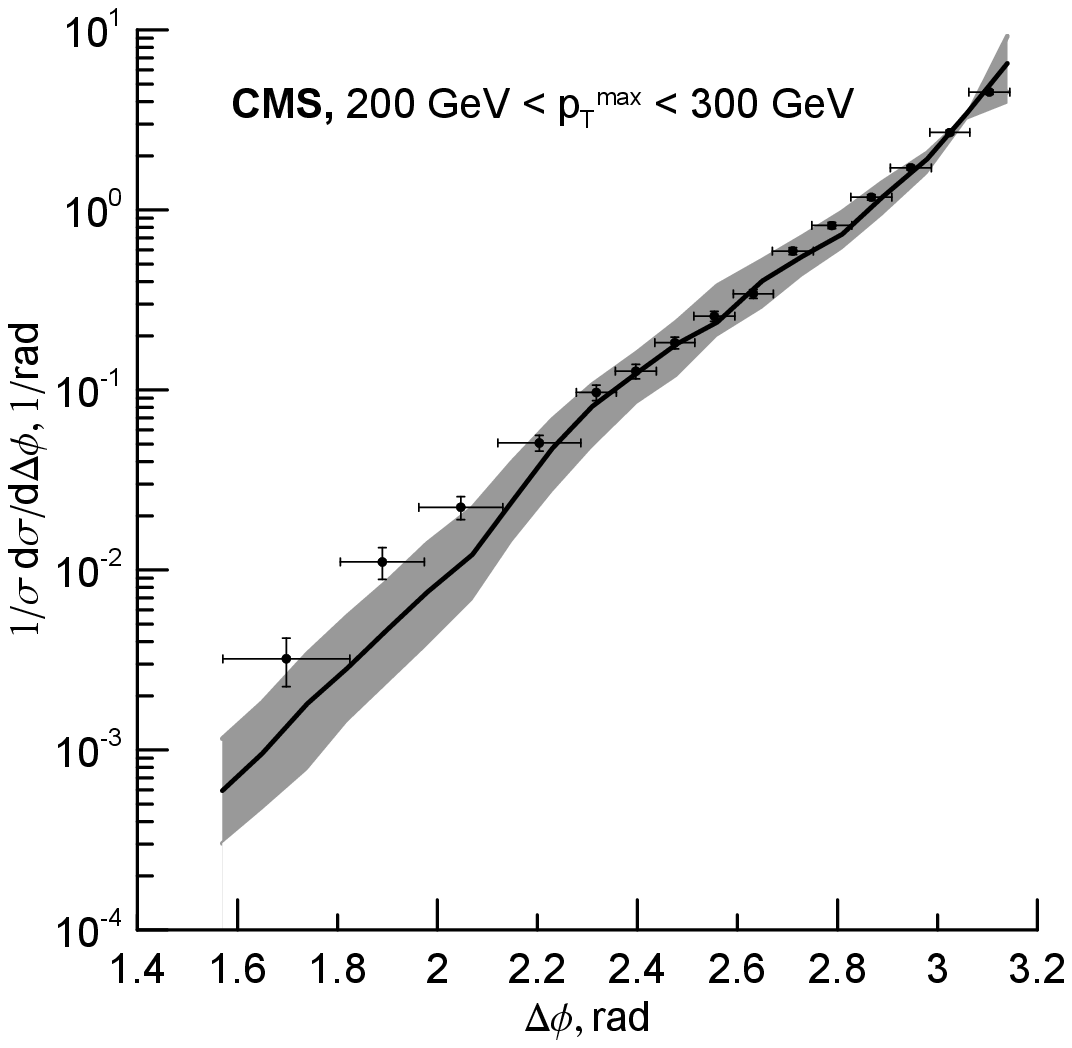}
\end{center}
\caption{Normalized $F(\Delta\phi)$ distributions in several
$p_T^{max}$ regions at the $\sqrt{S}=7$ TeV, $|y|<1.1$ and $p_T>30$
GeV. The data are from the CMS Collaboration \cite{CMS2}. The curve
corresponds to LO parton Reggeization approach.}\label{figCMS}
\end{figure}

The  CMS measurements \cite{CMS2}   of $F(\Delta\phi)$ distributions
for the two most energetic jets are shown in Fig.~\ref{figCMS}. A
comparison of the LO parton Reggeization approach predictions with
the data demonstrates a nice agreement in the region of
$\Delta\varphi\geq 3\pi/4$. As $\Delta\varphi$ decreases  from
$3\pi/4$ to $\pi/2$, the theoretical expectations tend to
underestimate data more and more, up to a factor of 5 at the
$80<p_T^{max}<110$ GeV and a factor 2 at the $200<p_T^{max}<300$
GeV. This difference follows from our theoretical approximation: we
take into account only two-jet production subprocesses in the QMRK,
like the $RR\to gg$. However, at $\Delta\varphi\simeq \pi/2$, the
contribution of three-jet production subprocess should be important.
One can find, the difference becomes smaller with the growing of
$p_T^{max}$, because the situation when transverse momentum of
leading jet is compensated by the one energetic jet in opposite
direction is more probable than the such compensation by two or more
jets.

\begin{figure}[ph]
\begin{center}
\includegraphics[width=.5\textwidth, clip=]{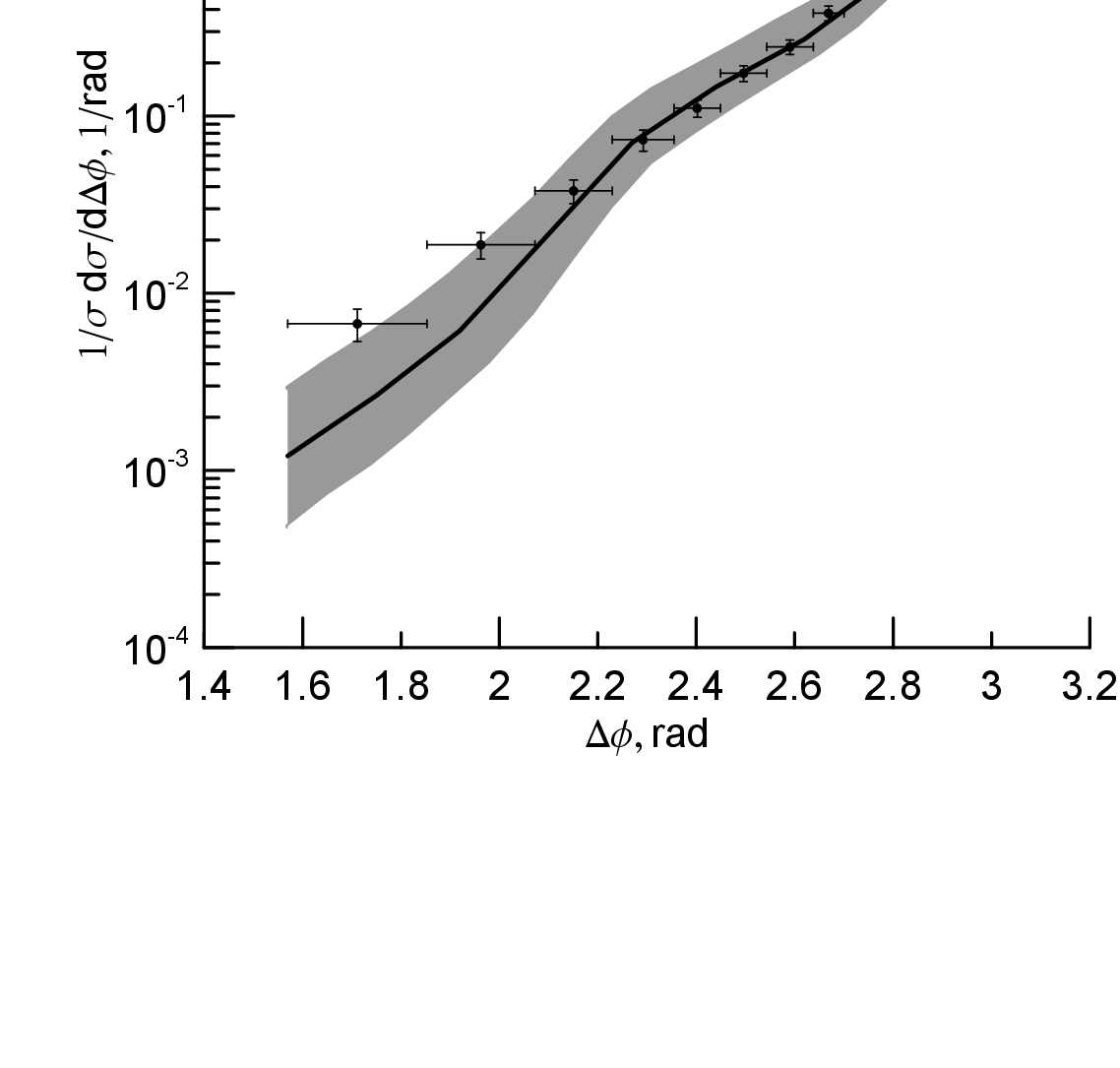}\includegraphics[width=.5\textwidth, clip=]{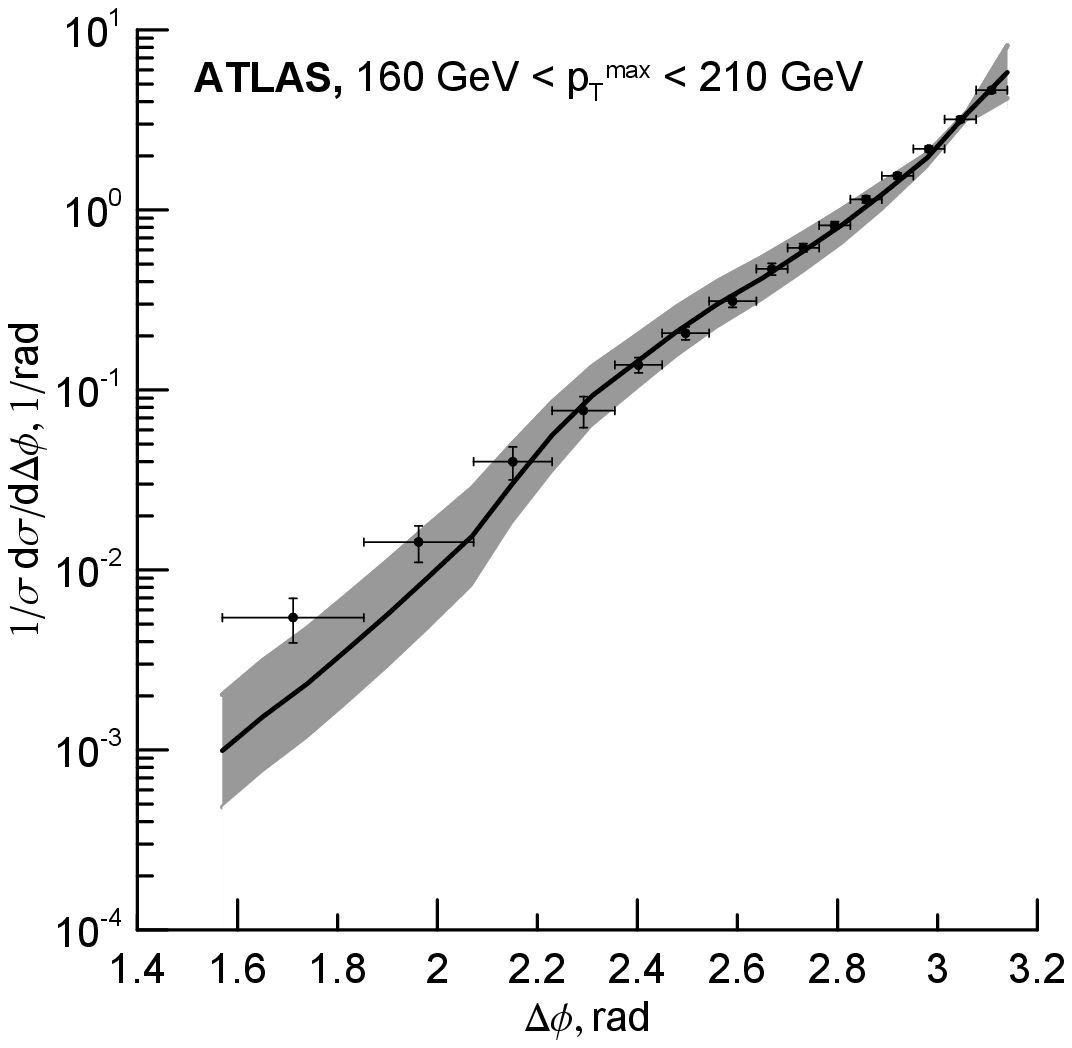}
\includegraphics[width=.5\textwidth, clip=]{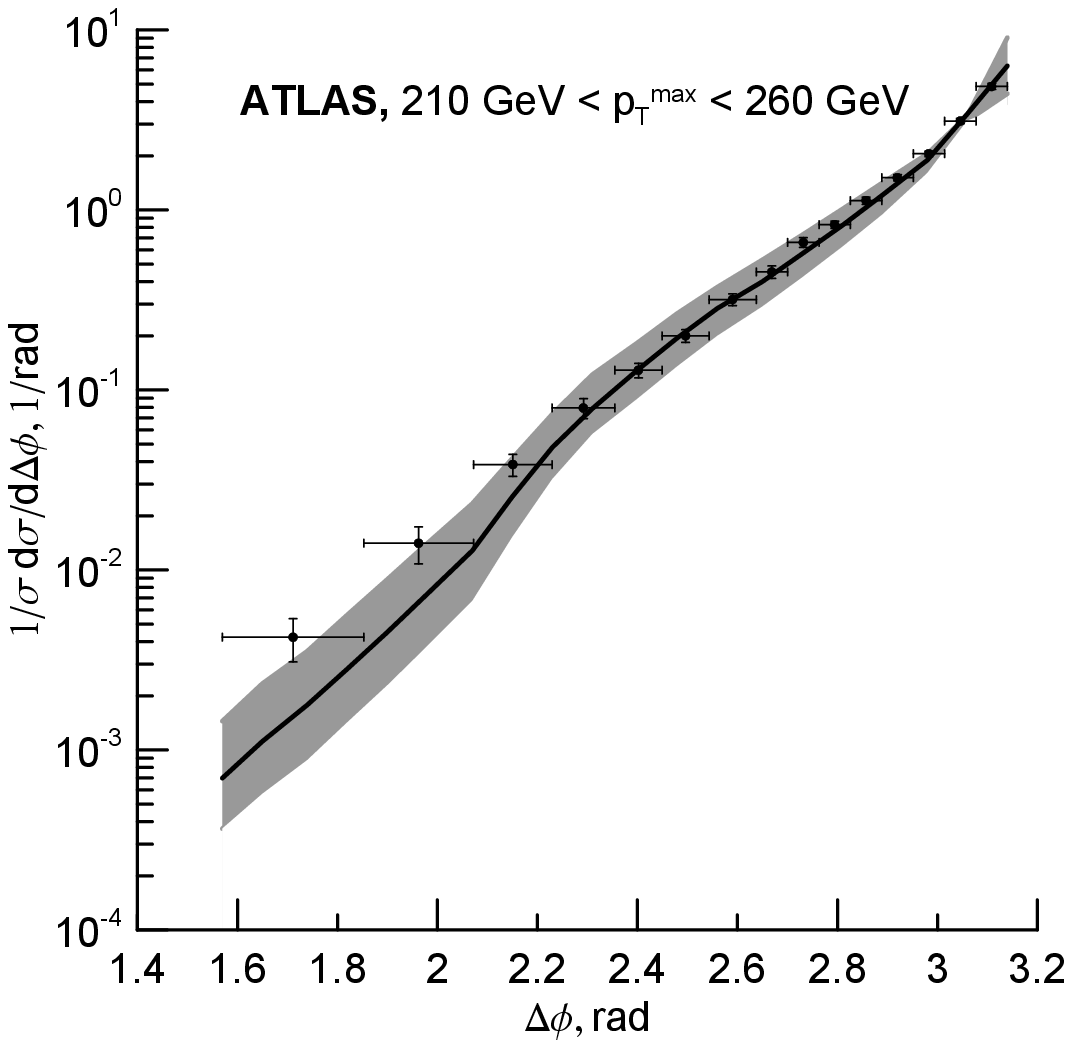}\includegraphics[width=.5\textwidth, clip=]{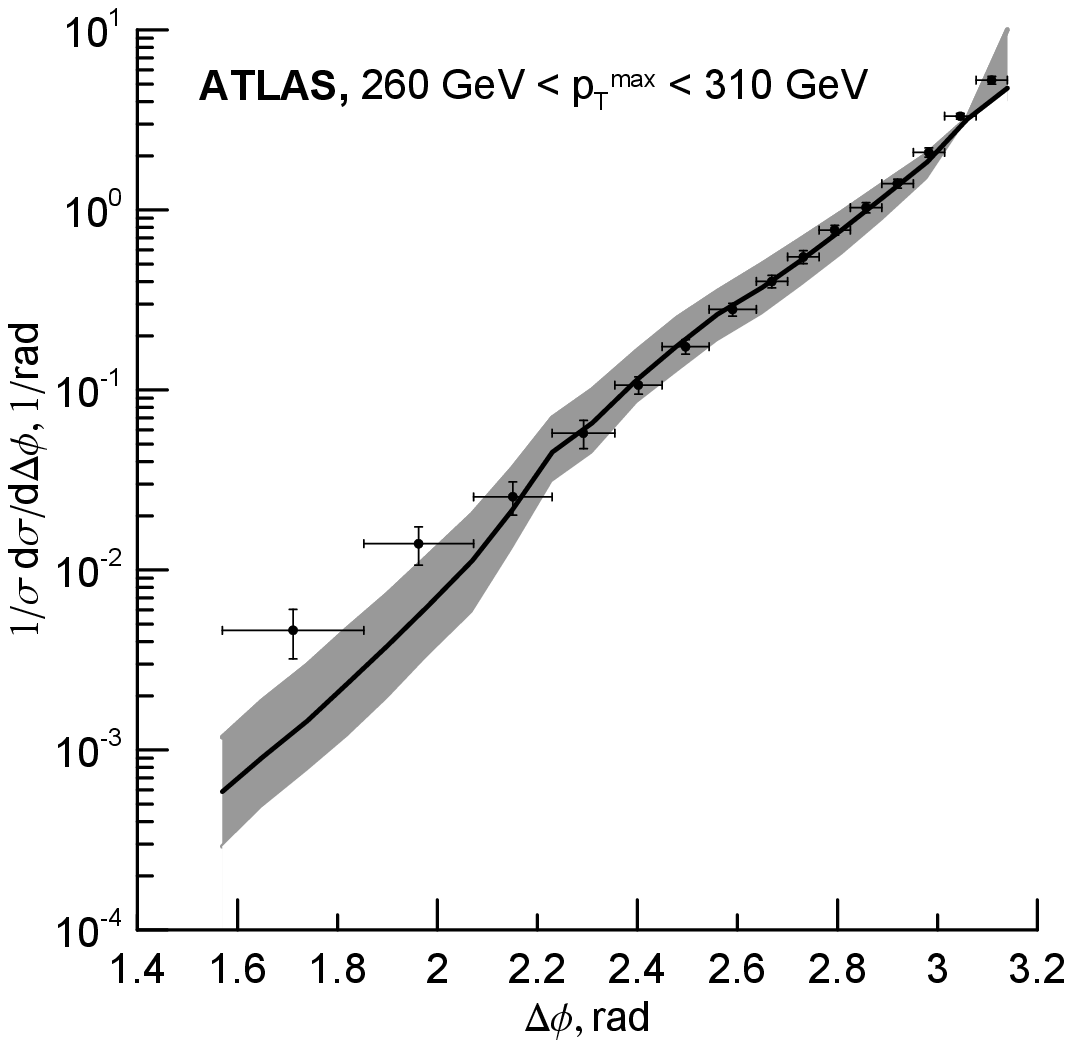}
\end{center}
\caption{Normalized $F(\Delta\phi)$ distributions in several
$p_T^{max}$ regions at the $\sqrt{S}=7$ TeV, $|y|<0.8$ and $p_T>100$
GeV. The data are from the ATLAS Collaboration \cite{ATLAS2}. The
curve corresponds to LO parton Reggeization approach.}\label{figATLAS}
\end{figure}

Our observations are confirmed in Fig.~\ref{figATLAS}, where the
data from the ATLAS Collaboration \cite{ATLAS2} are compared with
theoretical predictions. We see that the disagreements between our
predictions and ATLAS data are smaller than in the case of CMS data.
This fact can be explained by the difference of the choice of
low-$p_T$ cut for both Collaborations: CMS measurements have been
performed for jet production with $p_T>30$ GeV, and ATLAS
measurements -- with $p_T>110$ GeV.

\begin{figure}[ph]
\begin{center}
\includegraphics[width=.6\textwidth, clip=]{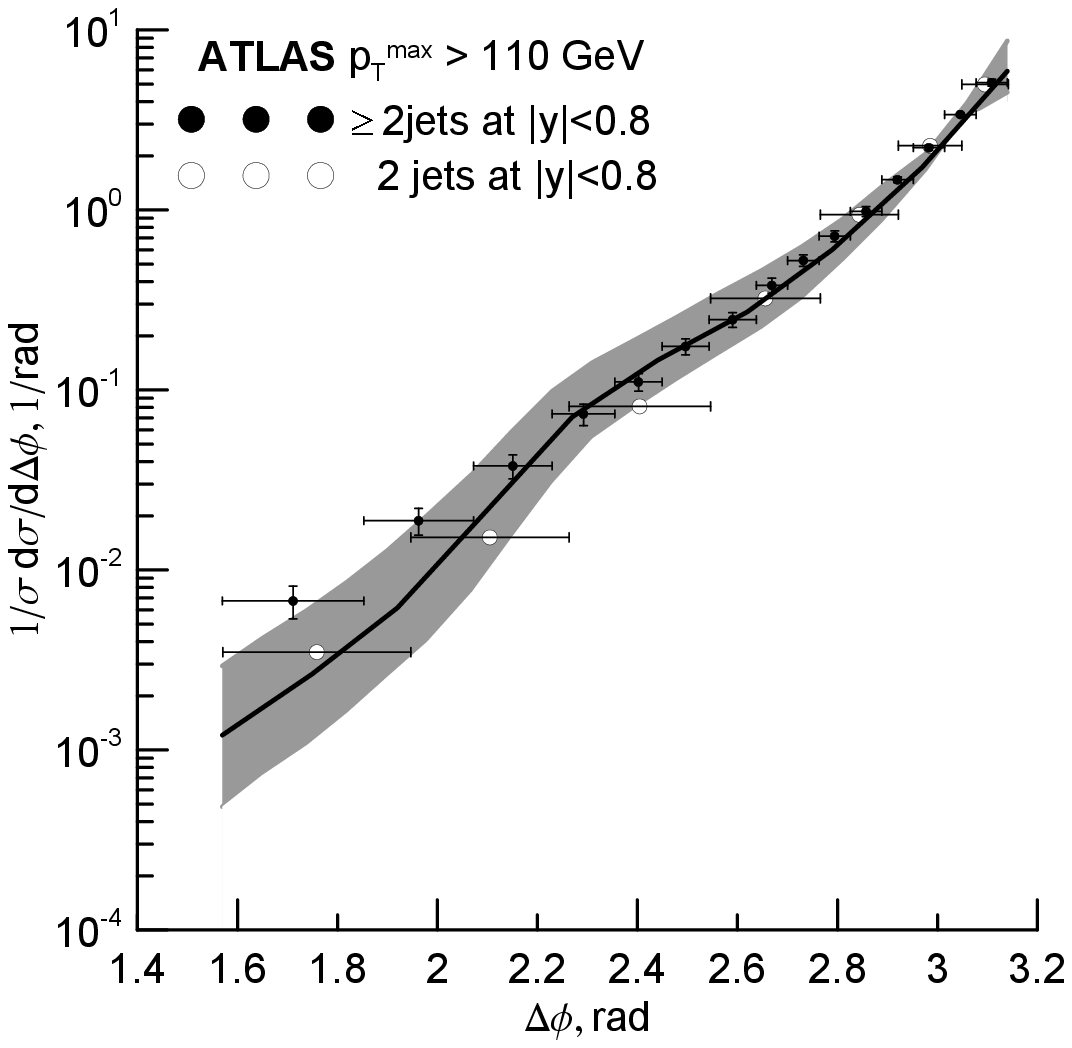}
\end{center}
\caption{Normalized $F(\Delta\phi)$ distribution for 2 (open circles)
and $\geq 2$ (black circles) jets with $p_T>100$ GeV, $|y|<0.8$,
$p_T^{max}>110$ GeV and $\sqrt{S}=7$ TeV. The data are from the
ATLAS Collaboration \cite{ATLAS2}. The curve corresponds to LO parton
Reggeization approach.}\label{ATLAS2to2}
\end{figure}

Certainly, the precise comparison of theoretical predictions in the
LO parton Reggeization approach should be made when we separate only
the two-jet production in the central rapidity region. In fact, the
data include a multijet production contribution. The ATLAS
Collaboration presents the $\Delta\varphi-$distributions for
different numbers of final-state jets (see Fig.~1 in
Ref.~\cite{ATLAS2}). Using these data, we can extract
$F(\Delta\varphi)$ for two-jet production only, as a difference
between number of events: $n(2)=n(\geq2)-n(\geq3)$ or
$\sigma(2)F(\Delta\varphi,2)=\sigma(\geq2)F(\Delta\varphi,\geq
2)-\sigma(\geq 3)F(\Delta\varphi,\geq3)$. The original ATLAS data
for $F(\Delta\varphi,\geq2)$ and extracted data for
$F(\Delta\varphi,2)$ are shown in Fig.~\ref{ATLAS2to2} for the
kinematic domain of $p_T^{max}>100$ GeV and $|y|<0.8$. As we see,
the theoretical prediction nicely agrees with data for
$F(\Delta\varphi,2)$ distribution. So, if the last one would be
extracted for different regions of $p_T^{max}$, we can make a more
precise comparison of our predictions with experimental data.

Summarizing the results of the present analysis for dijet production
at the LHC and our previous study of $b\bar b-$pair production at
the Tevatron and LHC \cite{bbTEV,bbLHC}, we find a strong difference
of theoretical interpretation of azimuthal decorrelation between
leading and subleading jets, in the collinear parton model and in
the parton Reggeization approach. In the first case, an azimuthal
decorrelation at different values of $\Delta\varphi$ is provided by
hard $2\to 3$ ($3\pi/4<\Delta\varphi<\pi$), $2\to 4$
($\pi/2<\Delta\varphi<3\pi/4$) partonic subprocesses,
correspondingly. The explanation of data in the region of
$\Delta\varphi<\pi/2$ in the framework of collinear parton model
becomes possible only because of initial-state radiation and
hadronization effects, and an agreement of theory expectations and
data is achieved using MC generators only.

Oppositely, in the parton Reggeization approach, the azimuthal
decorrelation is explained by the coherent parton emission during
the QCD evolution, which is described by the
transverse-momentum-dependent PDFs of Reggeized partons. Already in
the LO approximation, at the level of $2\to 2$ subprocesses with
Reggeized partons, we can account the main part of decorrelation
effect in dijet production, and we obtain a full description of data
in $b\bar b$ pair production.

\section{Conclusions}
This good description of dijet azimuthal decorrelations is achieved
in the LO parton Reggeization approach, without any ad hoc
adjustments of input parameters. By contrast, in the collinear
parton model, such a degree of agreement calls for NLO and NNLO
corrections and complementary initial-state radiation effects and ad
hoc nonperturbative transverse momenta of partons. In conclusion,
the parton Reggeization approach has once again proven to be a
powerful tool for the theoretical description of QCD processes
induced by Reggeized partons in the high-energy limit.
  As it was shown by  recent studies \cite{NLOPRA}, the one-loop calculations in this
  formalism lead the results for the NLO effective vertices, consistent with the earlier
  calculations based on unitarity relations \cite{oldNLOPRA}. These results open a possibility
  to extend the calculations of hard processes in the parton Reggeization approach to the complete NLO level.

\section*{Acknowledgements}

We are grateful to  B.~A.~Kniehl, E.~A.~Kuraev, L.~N.~Lipatov and
N.~N.~Nikolaev for useful discussions. The work of V.~S. was
supported by the Ministry for Science and Education of the Russian
Federation under Contract No.~14.B37.21.1182. The work of  M.~N. and
A.~S. was supported in part by the Russian Foundation for Basic
Research under Grant 12-02-31701-mol-a. The work of M.~N. is
supported also by the Grant of the Graduate Students Stipend Program
of the Dynasty Foundation. The work of A.~S. was also supported in
part by Michail Lomonosov Grant No. A/11/76586, jointly funded by
the German Academic Exchange Service DAAD and the Ministry of
Science and Education of the Russian Federation. V.S. and A.S. are
kindly grateful to B.~A.~Kniehl for the hospitality and the
computational resources provided during their visits. V.S. and A.S.
express gratitude to E. Kader for her help in technical aspects
during the preparation of this paper.



\newpage

\section*{Appendix}
1. $ RR \to g  g$
\begin{eqnarray} A&=&\frac{18}{a_1a_2b_1b_2
s^2t^2u^2t_1t_2},\nonumber\\
W_0&=& x_1 x_2 s^2 t u t_1 t_2 \bigl( x_1 x_2 (t u + t_1 t_2) + (a_1 b_2 + a_2 b_1) t u \bigr),\nonumber\\
W_1&=& x_1 x_2 s t_1 t_2 \biggl[t^2 u \biggl(a_1 b_2 (a_2 b_2+a_1 x_2)(t_1+t_2)-a_2b_1(a_1b_1t_1+a_2 b_2 t_2)+\nonumber\\
&+&\bigl(x_2(a_1^2b_2+a_2^2b_1)+a_1a_2(b_1-b_2)^2\bigr)u+x_1x_2 a_1 b_2t \biggr)\biggr]+\biggl[a_1\leftrightarrow a_2, b_1\leftrightarrow b_2, t\leftrightarrow u \biggr],\nonumber\\
W_2&=&a_1a_2b_1b_2tu\biggl(x_1^2x_2^2\bigl[2(t_1+t_2)\bigl(t^2u+t_1t_2(s+u-t)\bigr)+tu\bigl((t_1-t_2)^2+t(u+2t)\bigr)\bigr]+\nonumber\\
&+&x_1x_2tt_1t_2\bigl(4(x_1b_1+x_2a_2)(s+u)-(a_1b_1+a_2b_2)u\bigr)+\nonumber\\
&+&tu\bigl(x_1^2b_2(2x_2t-b_1t_1)t_1+x_2^2a_1(2x_1t-a_2t_2 )t_2\bigr)\biggr)+\biggl(a_1\leftrightarrow a_2, b_1\leftrightarrow b_2, t\leftrightarrow u \biggr),\nonumber\\
W_3&=&x_1x_2a_1a_2b_1b_2\biggl[t^2u\biggl(2a_1b_2\bigl(x_1x_2(t_1+t_2)(2t-u-s)-(x_1b_2t_1+x_2a_1t_2)(u+s)\bigr)+\nonumber\\
&+&\bigl[x_1t_1\bigl(2(a_1b_2^2+a_2b_1^2)+3x_1b_1b_2\bigr)+x_2t_2\bigl(2(a_1^2b_2+a_2^2b_1\bigr)+3a_1a_2x_2)\bigr] u+\nonumber\\
&+&4x_1x_2t\bigl((a_1b_2+a_2b_1)u+a_1b_2t\bigr)\biggr)\biggr]+\biggl[a_1\leftrightarrow a_2, b_1\leftrightarrow b_2, t\leftrightarrow u \biggr],\nonumber\\
W_4&=& x_1^2 x_2^2 a_1 a_2 b_1 b_2 \biggl[t \biggl(a_1a_2b_1b_2u(t_1+t_2)(t-u-s)+(a_1b_2+a_2b_1)^2 tu^2-\nonumber\\
&-&2a_1b_2t(s+u)(2a_2b_1u-a_1b_2s)\biggr)\biggr]+\biggl[a_1\leftrightarrow
a_2, b_1\leftrightarrow b_2, t\leftrightarrow u \biggr].
\end{eqnarray}

2. $ RR \to q  \bar q$
\begin{eqnarray}
A&=&\frac{S}{3s^2 t^2 u^2 t_1 t_2},\nonumber\\
W_0&=&18 x_1 x_2 s t^2 u^2 t_1 t_2,\nonumber\\
W_1&=&t u \biggl(-18 t u \bigl((x_1b_1t_1-a_1x_2t_2)+x_1x_2(u+t_2)\bigr)\bigl((x_1b_1t_1-a_1x_2t_2)-x_1x_2(t+t_1)\bigr)+\nonumber\\
&+&x_1x_2s\bigl[9\bigl((t-u)(a_1b_2-a_2 b_1)t_1 t_2 - x_1 x_2 stu \bigr)-x_1x_2s(t_1t_2-tu)\bigr]\biggr),\nonumber\\
W_2&=&x_1 x_2 t u \bigl[9\bigl(2(x_1 b_1 t_1-x_2 a_1 t_2)-x_1 x_2
(t+t_1-u-t_2)\bigr)\times\nonumber\\&\times&\bigl(2(a_2b_1-a_1b_2)tu-(a_1b_2t-a_2b_1u)s\bigr)-
7x_1x_2s^2(a_2b_1u+a_1b_2t)\bigr],\nonumber\\
W_3&=&-x_1^2x_2^2\bigl[18tu(a_2b_1-a_1b_2)\bigl((a_2b_1-a_1b_2)tu+(a_2b_1u-a_1b_2t)s\bigr)+
\nonumber\\
&+&2 s^2(4a_2^2b_1^2u^2+4a_1^2b_2^2t^2-a_1a_2b_1b_2tu)\bigr],\nonumber\\
W_4&=&0.
\end{eqnarray}

3. $Q R \to q g$
\begin{eqnarray}
A&=&-\frac{8 x_1}{9 a_2 b_1 b_2  s
 t ^2  u^2 t_2},\nonumber\\
W_0&=&-9x_2t_2 s t u\bigl(x_2t_1t_2+(x_2+b_2)
t u\bigr),\nonumber\\
W_1&=& t u\biggl(-a_2b_1b_2 t u\bigl[8\bigl(b_1( s+ t)-b_2 u\bigr)+x_2 s\bigr]+\nonumber\\
&+&\bigl[9x_2^2\bigl(a_2(b_1-b_2) s^2+(a_2b_1+a_1b_2) s t-a_1b_2 t u\bigr)+\nonumber\\
&+&9x_2a_2 u\bigl((b_1^2-2b_1b_2-b_2^2) s-b_1b_2 u\bigr)+\nonumber\\
&+&x_2a_2b_1 t\bigl((b_1-b_2)s+ b_1t\bigr)+2a_2b_1 t u(b_1^2+b_1b_2+4b_2^2)\bigr]t_2+\nonumber\\
&+&x_2\bigl((a_2b_1 t+9a_2x_2 s)(b_1-b_2)+9b_1b_2 u(a_1-a_2)\bigr)t_2^2\biggr),\nonumber\\
W_2&=&a_2b_1b_2x_2 t u\biggl(9\bigl(a_1 t(x_2 s+b_2 u)-a_2 u(x_2
s-b_2 u)-2a_2b_1 u(
t+ s)\bigr)+\nonumber\\
&+&b_1 t(a_1 t+a_2 u)-2a_1b_2 t( s+
u)+\nonumber\\
&+&\biggl[9\bigl(x_1x_2( s+ u)+2a_1b_1
u+x_2a_1( s-3 u)\bigr)+(b_1x_1-2a_1b_2) t\bigr]t_2\biggr),\nonumber\\
W_3&=&a_2 b_1 b_2 x_2^2 \biggl(9\bigl[ s\bigl(a_1^2b_2 t^2-2
a_2^2b_1 u^2+a_1a_2 u t(b_1-b_2)\bigr)+a_2 t
u^2(a_1b_2-a_2b_1)\bigr]-\nonumber\\
&-&a_1t^2\bigl( u(a_1b_2-a_2b_1)+a_1b_2
s\bigr)\biggr),\nonumber\\
W_4&=&0.
\end{eqnarray}

\newpage
4. $Q Q \to q  q$

\begin{eqnarray}
A&=&\frac{64 x_1x_2}{27 a_1a_2b_1b_2 t^2
u^2S},\nonumber\\
W_0&=&-stut_1t_2,\nonumber\\
W_1&=&t^2 u^2\bigl[6(a_1b_2+a_2b_1)-a_1b_1-a_2b_2\bigr]+ s t u(a_1b_2 t+a_2b_1 u)-\nonumber\\
&-& t u\bigl[t_1(a_1b_1 t+a_2b_2 u)+t_2(a_1b_1 u+a_2b_2 t)+t_1t_2(a_1-a_2)(b_1-b_2)\bigr],\nonumber\\
W_2&=&- t u\bigl[x_2^2a_1a_2t_1+x_1^2b_1b_2t_2+a_1b_2
t(a_1b_1+x_1b_2-5a_2b_1)+ a_2b_1
u(a_1b_1+x_2a_2-5a_1b_2)\bigr],\nonumber\\
W_3&=&3a_1a_2b_1b_2(a_1b_2 t^2+a_2b_1 u^2),\nonumber\\
W_4&=&0.
\end{eqnarray}

5. $Q Q' \to q q'$
\begin{equation}
A=\frac{64 x_1x_2}{9 a_2b_1 t^2},\qquad W_0=2 t^2,\quad W_1=2 a_2
b_1 t,\quad W_2=a_2^2b_1^2, \quad W_3=0,\quad W_4=0.
\end{equation}

6. $Q \bar Q \to q \bar q$
 \begin{eqnarray}
A&=&\frac{64}{27 x_1x_2a_2b_1 s^2 t^2},\nonumber\\
W_0&=& x_1x_2 s t \bigl[t_1 t(3a_2b_1-x_1b_2)+t_2 t
(3a_2b_1-x_2a_1)+
t_1t_2(x_2a_2-x_1b_2)-x_1x_2 t^2+\nonumber\\
&+& s t\bigl(6(a_1b_1+a_2b_2)+5(2a_2b_1+a_1b_2)\bigr)\bigr],\nonumber\\
W_1&=& t\bigl[t_1x_1a_2b_2\bigl(6b_1 t(a_2b_1-a_1b_2)-
x_2 s(x_1b_1+a_2b_1-a_1b_2)\bigr)+\nonumber\\
&+&t_2x_2a_1b_1\bigl(6a_2 t(a_2b_1-a_1b_2)-
x_1 s(x_2a_2+a_2b_1-a_1b_2)\bigr)+\nonumber\\
&+&6x_1x_2a_2b_1(a_2b_1-a_1b_2) t^2+x_1x_2a_2b_1 s^2(a_2b_1-a_1b_2+6x_1x_2)+\nonumber\\
&+&x_1x_2 s t\bigl((a_1b_2-a_2b_1)^2+a_1b_2(a_1b_1+a_2b_2)-
2a_2b_1(2a_2b_1+x_1b_2+x_2a_1)\bigr)\bigr],\nonumber\\
W_2&=&x_1x_2a_2b_1\bigl({6 t^2}(a_2b_1-a_1b_2)^2+3x_1x_2a_2b_1 s^2+
s
t(x_1b_1+x_2a_2)(a_1b_2-a_2b_1)\bigr),\nonumber\\
W_3&=&W_4=0.
\end{eqnarray}

7. $Q \bar Q \to q' \bar q'$
\begin{eqnarray}
A&=&\frac{64}{9x_1x_2 s^2},\nonumber\\
W_0&=&-x_1x_2  s(  t+  u),\nonumber\\
W_1&=&2\bigl(  a_1^2b_1b_2(t_1+  u)+a_1^2b_2^2
u+a_1a_2b_1^2(  t+t_2)+a_1a_2b_1b_2(t_1+t_2-  s)+\nonumber\\
&+& a_1a_2b_2^2(t_2+  u)+a_2^2b_1^2  t+a_2^2b_1b_2(t_1+
t)\bigr),\nonumber\\
W_2&=& 2x_1x_2(a_2b_1-a_1b_2)^2,\nonumber \\
W_3&=&W_4=0.\end{eqnarray}

8. $Q  \bar Q \to g  g$
\begin{eqnarray}
A&=&\frac{1}{27 x_1x_2 S a_1a_2b_1 b_2  s^2 t^2 u^2 },\nonumber\\
W_0&=&2x_1x_2s^2tu\bigl[\bigl(x_1x_2-9(a_1b_2+a_2b_1)\bigr)(t+u)tu-x_1x_2t_1t_2(t_1+t_2)\bigr]+
\bigl[a_1\leftrightarrow a_2, b_1\leftrightarrow b_2, t\leftrightarrow u \bigr],\nonumber\\
W_1&=&-2x_1x_2stu\biggl(x_1x_2(a_1-a_2)(b_1-b_2)(t_1+t_2)t_1t_2+
2t\bigl[x_1x_2(x_1b_2t_2^2+x_2a_1t_1^2)+\nonumber\\
&+& t_1t_2\bigl((x_1x_2+x_1b_2+x_2a_1)(a_1b_1+a_2b_2)+
(8a_2b_1-a_1b_2)(a_2b_1-a_1b_2)\bigr)\bigr]+\nonumber\\
&+&x_1x_2\bigl(2x_1x_2-17(a_1b_2+a_2b_1)\bigr)(t_1+t_2)tu+2\bigl(x_1^2(x_2+b_1)(b_2-8b_1)+\nonumber\\
&+&
x_2^2a_1(a_2-8a_1)+a_1b_1(25x_1x_2-72a_2b_2)\bigr)t^2u+2t^2(x_1x_2 - 9a_2b_1)(x_1b_2t_2 + x_2a_1t_1)\biggr)+\nonumber\\
&+&\biggl(a_1\leftrightarrow a_2, b_1\leftrightarrow b_2, t\leftrightarrow u \biggr),\nonumber\\
W_2&=&\biggl[-2tu\biggl(-x_1^2x_2^4a_1a_2(t_1+t_2)^2t_1+2x_1x_2a_1 tt_1(t_1+t_2)\bigl[x_2^2\bigl(9a_1^2b_1-x_1(x_1x_2+a_1b_2)\bigr)+\nonumber\\
&+&x_1(8b_1-b_2)\bigl(x_2^2(a_2-a_1)+a_2b_2^2\bigr)\bigr]+\nonumber\\&+&x_1tut_1\bigl(36x_1a_1a_2b_2^3(b_2-2x_2)+5a_1a_2x_1x_2^2b_2^2-x_2^4a_2(x_1+a_1)(2a_1-7a_2)+\nonumber\\
&+&9x_2^2b_2(x_1^3b_2-2x_2a_1^3+4x_1^2x_2a_1)-4x_1x_2^3b_2(a_1^2+4x_1^2)\bigr)+2x_1x_2a_1t^2t_1\times\nonumber\\
&\times&\bigl[9a_2\bigl(b_2^2(3x_1b_1-2x_2a_1)-x_2^2b_2(x_1-3a_1)+x_2^3a_2\bigr)-2x_1x_2b_2(a_1x_2+a_2b_2)-x_1x_2^3a_2\bigr]+\nonumber\\
&+&x_1x_2t^2u\bigl(a_1^3b_2(7b_2^2+10b_1b_2-4b_1^2)+a_1^2a_2b_1(39b_2^2+30b_1b_2-2b_1^2)+\nonumber\\
&+&14a_1a_2^2b_1^2(x_2+2b_2)+8a_2^3b_1^3\bigr)+x_1x_2a_1b_2t^3\bigl(18a_2b_1(a_1x_2+x_1b_2)-x_1x_2(a_1b_1+x_1b_2)\bigr)\biggr)+\nonumber\\&+&\biggl(a_1\leftrightarrow b_2, a_2\leftrightarrow b_1, x_1\leftrightarrow x_2, t_1\leftrightarrow t_2 \biggr)\biggr]+\biggl[a_1\leftrightarrow a_2, b_1\leftrightarrow b_2, t\leftrightarrow u \biggr],\nonumber\\
W_3&=&-4x_1x_2a_1a_2b_1b_2\biggl(t^2\bigl[8x_1x_2a_1b_2(s+u)^2-u(s+u)\bigl((9a_2b_1+7a_1b_2)(a_1b_1+a_2b_2)+\nonumber\\
&+&2a_1b_2(17a_2b_1-a_1b_2)\bigr)+2a_1u^2\bigl(4b_1(a_1b_2+a_2b_1)+b_2(13a_1b_2-5a_2b_1)\bigr)\bigr]\biggr)+\nonumber\\
&+&\biggl(a_1\leftrightarrow a_2, b_1\leftrightarrow b_2, t\leftrightarrow u \biggr),\nonumber\\
W_4&=&0.
\end{eqnarray}

\newpage

\begin{center}
\textbf{Erratum: \\ M.A.~Nefedov, V.A.~Saleev, A.V.~Shipilova, Dijet
azimuthal decorrelations at the LHC in the parton Reggeization
approach, Phys. Rev. D87, 094030 (2013)}
\end{center}

The following corrections should be made to the expressions for
squared matrix elements of $2\to 2$ processes in parton Reggeization
approach, which has been published in the Appendix of Phys. Rev.
D87, 094030 (2013).
\begin{enumerate}
\item For the squared matrix element 3, the factor $a_1$ has been lost in the last term of $W_3$ in the published version. The correct
term has a form: $a_1t^2\bigl( u(a_1b_2-a_2b_1)+a_1b_2 s\bigr)$.
\item In the squared matrix element 4 the whole term $(\pi^2 \alpha_s^2)\cdot\dfrac{64 x_1x_2}{27 a_1a_2b_1b_2 t^2
u^2}\cdot (-stut_1t_2/S)$ should be added to the squared amplitude,
presented in the journal. Restoration of the structure of
decomposition of the squared amplitude in powers of $S$ requires the
change of $A$ and $W_i$ coefficients.
\item While interpreting the published expression for the squared matrix element 8, the reader should note, that the replacements
$a_1 \leftrightarrow b_2$ and $a_2 \leftrightarrow b_1$ in the
square braces of $W_2$ should be accompanied by the replacement $x_1
\leftrightarrow x_2$, since $x_1=a_1+a_2$ and $x_2=b_1+b_2$.
\end{enumerate}

  The given misprints in the expressions appeared only at the stage of preparation to the publication.
  The equivalent but less compact form of the squared amplitudes has been used in the numerical calculations, therefore,   the published numerical results are unaffected.

Recently [K. Kutak, R. Maciula, M. Serino, A. Szczurek, A. van
Hameren, "Four-jet production in single- and double-parton
scattering within high-energy factorization", JHEP 1604 (2016) 175],
it has been performed a comparison of the cross sections for dijet
production obtained both with the QCD amplitudes from the parton
Reggeization approach and those from the AVHLIB [A. van Hameren,
"BCFW recursion for off-shell gluons," JHEP, 07(2014), 138, 1404.
7818; A. van Hameren and M. Serino, "BCFW recursion for TMD parton
scattering," JHEP, 07(2015), 010, 1504.00315].  The perfect
agreement has been found. The consistency between amplitudes
computed in different approaches is a non-trivial check of the
methods employed.

The current version of the Appendix contains the corrected
expressions for the coefficients of the decomposition of the squared
amplitudes.

We wish to thank Krzysztof Kutak, Rafal Maciula, Mirko Serino,
Antoni Szczurek, and Andreas van Hameren for useful discussion of the
obtained results on analytic formula for matrix elements.

\end{document}